\pdfoutput=1

\documentclass[10pt]{article}
\usepackage{amsmath}
\usepackage{amssymb}
\usepackage{float}
\usepackage[labelfont={sf,bf}]{caption}
\usepackage{graphicx}
\usepackage{hyperref}
\usepackage{enumitem}
\usepackage{tabularx}
\usepackage{cite}
\usepackage{wasysym}
\usepackage{authblk}
\usepackage{abstract}

\usepackage{sectsty}
\allsectionsfont{\normalfont \sffamily \bfseries}
\usepackage{geometry}
\begin{document}

\title{\textbf{\textsf{Upgrade of the ultracold neutron source at the pulsed reactor TRIGA Mainz}}}
\author[1]{J. Kahlenberg}
\author[3,4]{D. Ries}
\author[1]{K. U. Ross}
\author[2]{C. Siemensen}
\author[1]{M. Beck}
\author[2]{C. Geppert}
\author[1]{W. Heil\thanks{Email: \href{mailto:wheil@uni-mainz.de}{wheil@uni-mainz.de}}}
\author[3,4]{N. Hild}
\author[1]{J. Karch}
\author[2]{S. Karpuk}
\author[1]{F. Kories}
\author[1]{M. Kretschmer}
\author[3]{B. Lauss}
\author[2]{T. Reich}
\author[2]{Y. Sobolev}
\author[2]{N. Trautmann}
\affil[1]{Institute of Physics, Johannes Gutenberg University, Mainz, Germany}
\affil[2]{Institute of Nuclear Chemistry, Johannes Gutenberg University, Mainz, Germany}
\affil[3]{Laboratory for Particle Physics, Paul Scherrer Institute (PSI), Villigen, Switzerland}
\affil[4]{Institute for Particle Physics, ETH Z\"u{}rich, Switzerland}
\date{\today}
\maketitle

\begin{abstract}
\noindent
\normalsize
The performance of the upgraded solid deuterium ultracold neutron source at the pulsed reactor TRIGA Mainz is described. The current configuration stage comprises the installation of a He liquefier to run UCN experiments over long-term periods, the use of stainless steel neutron guides with improved transmission as well as sputter-coated non-magnetic $^{58}$NiMo alloy at the inside walls of the thermal bridge and the converter cup. The UCN yield was measured in a `standard' UCN storage bottle (stainless steel) with a volume of 32 litres outside the biological shield at the experimental area yielding UCN densities of 8.5 /cm$^3$; an increase by a factor of 3.5 compared to the former setup. The measured UCN storage curve is in good agreement with the predictions from a Monte Carlo simulation developed to model the source. The growth and formation of the solid deuterium converter during freeze-out are affected by the ortho/para ratio of the H$_2$ premoderator. 
\end{abstract}

\vspace{1cm}

\section{Introduction}
\label{intro}
Neutrons are termed ``ultracold neutrons'' (UCN) when they are slow enough to be confined in traps. These traps may be formed by i) the strong interaction (Fermi) potential $V_{\text{F}}$ in suitable material bottles ($V_{\text{F}}$($^{58}$Ni) = 358 neV) \cite{Ref1, Ref2}, ii) magnetic potential walls (60 neV/Tesla), and iii) the potential energy in the Earth's gravitational field (100 neV/m) or by a combination of magnetic and gravitational trapping. Stored UCN behave like an ideal gas at a temperature in the millikelvin region and may be observed for very long times, in principle only limited by their lifetime. Hence, storage times of several minutes can be obtained, compared to observation times of only milliseconds in typical beam experiments. The resulting storage experiments enable high-precision measurements of the properties of the free neutron such as its lifetime \cite{Ref3, Ref4, Ref5, Ref5a, Ref5b, Ref5c} or the search for a finite electric dipole moment (EDM) \cite{Ref6, Ref7, Ref8, Ref9}, probably the best-known ``classical'' experiment using UCN. The scope of UCN experiments has been extended and includes measurements of neutron decay correlations \cite{Ref10, Ref11, Ref12}, searches for a finite electrical charge of the neutron \cite{Ref13, Ref14, Ref15, Ref15a}, investigations of gravitationally bound quantum states of neutrons \cite{Ref16, Ref17, Ref18, Ref19}, and tests of the weak equivalence principle for the neutron \cite{Ref20, Ref21, Ref22} or the theory of neutron diffraction \cite{Ref23}. \par
The limitation in all these measurements is the maximum attainable density of UCN. Thermal sources of UCN rely on extracting very low energy neutrons from the Max\-wellian tail of the thermal energy distribution in nuclear reactor-driven moderators. The final step to the ultracold state can be made by reflecting these neutrons off receding turbine blades as performed at the existing source at the Institute Laue-Langevin \cite{Ref24}. New UCN sources are presently developed worldwide based on the principle of superthermal UCN production, i.e., neutrons not in thermal equilibrium with the converter, using cryo-converters made of solid deuterium or superfluid helium. The idea that solid deuterium (sD$_2$) can be used as a superthermal UCN source goes back to a paper by Golub and B\"oning \cite{Ref25}. Pokotilovski pointed out the advantages of UCN production in sD$_2$ at pulsed neutron sources \cite{Ref26},  the use of spallation as a pulsed source for neutrons was suggested by the Gatchina group \cite{Ref27}. The first realisation of a Pokotilovski-type source was the Los Alamos source \cite{Ref27a, Ref27b}. Superthermal UCN sources are now in operation or under construction at different facilities worldwide, such as PSI (Villigen) \cite{Ref28, Ref29, Ref30, Ref31}, LANCSE (Los Alamos) \cite{Ref32}, ILL (Grenoble) \cite{Ref33, Ref34}, RCNP (Osaka University) \cite{Ref35}, PULSTAR reactor (NC State University) \cite{Ref36}, FRM-II (Garching) \cite{Ref37}, and TRIUMF (Vancouver) \cite{Ref38}. \par
Low-power reactors, such as the TRIGA Mainz, are competitive due to the possibility to pulse the reactor every five minutes while producing a high density of UCN in the pulse that ideally meets the requirements of storage experiments, in which the trap must be filled with a similar frequency. The UCN source moderates and converts thermalised neutrons to UCN in a sD$_2$ volume inside of a guide system. The neutrons are transported to an experimental area outside the biological shield. The technical aspects of the source are discussed in Ref. \cite{Ref39}.  Its performance after the first stage of construction is described in detail in Ref. \cite{Ref40}. Here, we report on results that have been obtained with the upgraded UCN source at the radial beamport D of the TRIGA Mainz.

\section{The TRIGA Mainz UCN source}
\label{sec:1}
For pulse mode operation, the reactor is brought to criticality at a low steady-state power (50 W$_{\text{th}}$). Then, the pulse rod is shot out of the reactor core by compressed air. Due to this sudden insertion of excess reactivity, the power rises sharply with a reactor period of only a few milliseconds. At the TRIGA Mainz reactor, a maximum peak power of 250 MW$_{\text{th}}$ in the pulse mode is obtained with a pulse width at half maximum (FWHM) of about 30 ms resulting in a maximum pulse energy of $\approx$ 10 MWs. The maximum allowed pulse rate ($R_\text{pulse}$) is 12 per hour. In the described experiments, the pulse rate was chosen to be $R_{\text{pulse}}$ = 5/h. \par

\begin{figure}
\centering
\resizebox{0.7\textwidth}{!}{
  \includegraphics{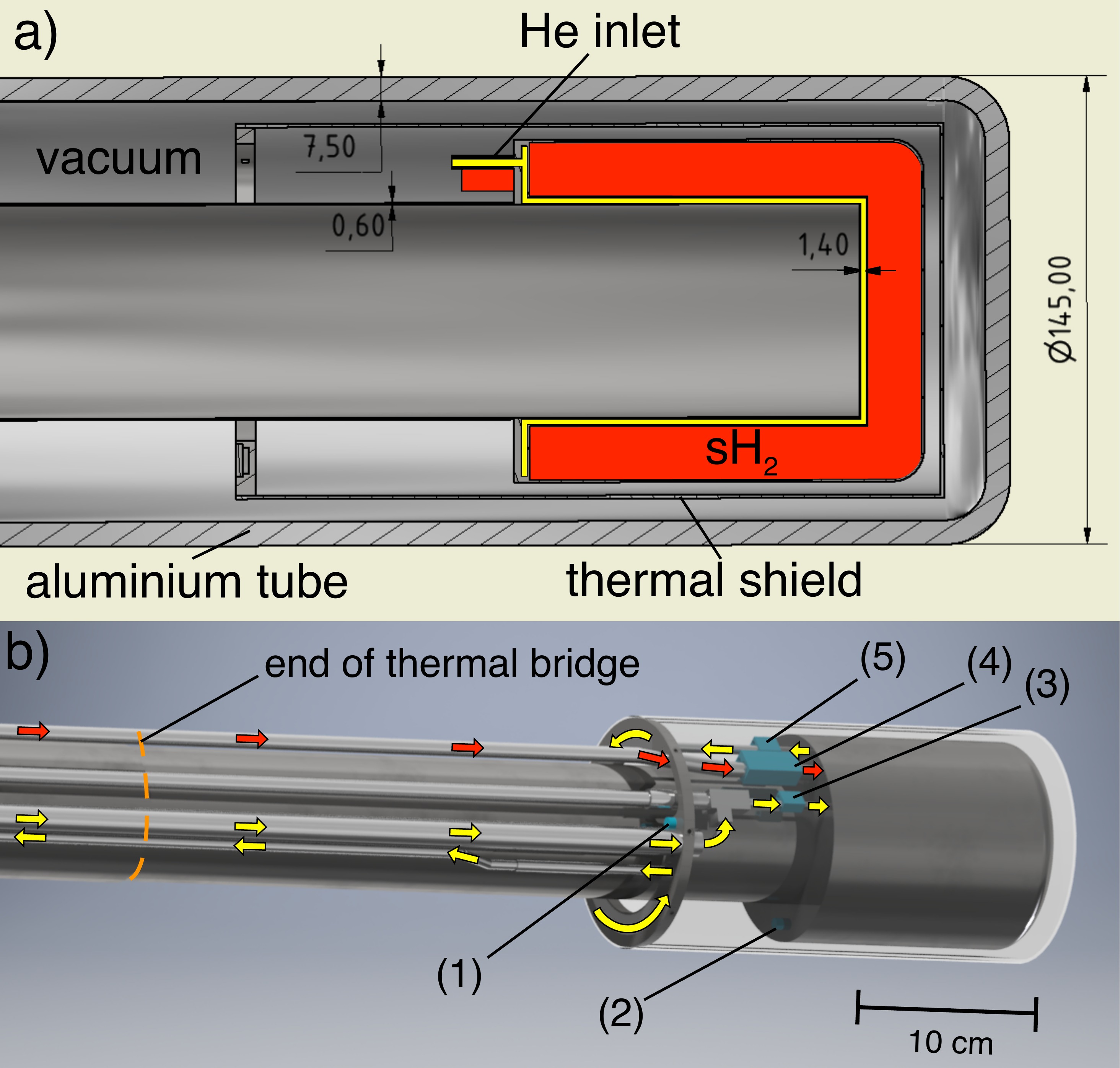}
}
\caption{\textbf{a)} Blow-up drawing of the front part of the in-pile cryostat showing the thermal bridge as part of the UCN guide which terminates in the cup-shaped, double-walled nose (converter cup) with cooling pipes for liquid helium. A third wall around the actual converter cup encloses the premoderator volume (red area). \; \; \; \; \; \; \; \; \; \; \; \; \; \; \; \; \; \; \; \; \; \; \; \;  \; \; \; \; \;  \; \; \;  \; \; \; \; \; \; \; \; \; \; \; \; \; \; \; \; \; \; \; \; \; \; \; \; \; \textbf{b)} Arrangement for He coolant flow (flow direction indicated by yellow arrows), H$_2$ feed pipe (red arrows), location of temperature sensors (Lake Shore Cernox$^{\text{TM}}$) with their respective temperature values during steady state operation: (1) Sensor housing thermal shield ($\approx$ 50 K), (2) sensor housing nose ($\approx$ 5 K), (3) sensor housing He inlet ($\approx$ 5 K), (4) sensor housing H$_2$ inlet ($\approx$ 8 K), and (5) sensor housing He outlet ($\approx$ 6 K). The relatively strong heat load on the sensors' stainless steel housing prevents the monitoring of the actual temperatures of the cryogenic solids shortly after the pulse.}
\label{fig:1}
\end{figure}

A schematic drawing of the UCN source at beamport D which is horizontally directed straight at the reactor core can be found in Ref. \cite{Ref40}. Therefore, only a short description is given: The cryostat consists of a vessel outside the reactor shielding and the in-pile part. The in-pile part of the cryostat houses the UCN guide, an electropolished stainless steel tube with an inner diameter $\diameter_{\text{ID}}$ = 66 mm and a total length of $\approx$ 4.4 m. At the position of the vertical cryostat, the in-pile part of the UCN guide is terminated by an AlMg3 separation foil with a thickness of 100 $\mu$m in order to provide clean vacuum conditions in the part of the UCN guide which is in direct contact with the deuterium gas. In Fig. \ref{fig:1}, a detailed drawing of the critical components of the source, i.e., moderator and premoderator, is shown along with operational parameters. The guide tube, which is kept at room temperature, is connected to the nose via a thermal bridge, i.e., a section of about 50 cm length, where the wall thickness of the stainless-steel tube is reduced to 0.6 mm to reduce heat transfer from the hot to the cold part. It terminates in the cup-shaped, double-walled nose with cooling pipes for liquid helium. Thus, it is guaranteed that all the D$_2$ gas is frozen out in the nose (total length 10 cm), which can contain  up to 160 cm$^3$ of sD$_2$ ($\approx$ 8 mol). In order to convert para-D$_2$ into ortho-D$_2$, we have a separate cryogenic cell with OXISORB \cite{Ref41} as catalyst. Hence, D$_2$ is always prepared and filled in the $>$ 98 \% ortho state\footnote{Using a Raman spectrometer, the para-D$_2$ concentration was routinely measured before and after each measurement run, while no change was found. Hydrogen contamination, a source of nuclear absorption in sD$_2$, has also been investigated by Raman scattering. Within the precision of the measurements (0.5 \%), no hydrogen content was found.}. A third wall around the actual converter cup encloses the premoderator volume of $\approx$ 620 cm$^3$ which corresponds to a premoderator thickness of $\approx$ 17 mm when the cup-shaped container is filled with solid H$_2$. At beamport D, optimum operation conditions at the highest possible thermal neutron flux with still tolerable heat load were found using a bismuth/graphite stopper which was put to the very end of beamport D just in front of the reactor core. During the measurements, the in-pile cryostat either touched the graphite/bismuth stopper directly (Pos. I) or was pulled back by 3 cm (Pos. II). The thermal neutron fluence and thus the expected UCN yield differs by $\approx$ 10 \% in the two positions of the source as can be extracted from Fig. 3 in Ref. \cite{Ref40}. \par When the UCN exit the sD$_2$ into the vacuum of the guide, they receive a potential boost of $V_{\text{F}}$(sD$_2$) = 105 neV \cite{Ref42}. Hence the UCN spectrum starts at this energy. \par
The upgrade of the source comprises
\begin{enumerate}[label=\alph*)]
\item{The He liquefier (TCF10, Linde) together with a permanently installed 1500 L dewar serving as He buffer which replaces the 250 L dewars having ensured He supply from a neighboring facility in the past.  With a maximum capacity of 14 L/h, the liquefier covers the consumption of liquid helium necessary for cooling the converter to its work temperature of $\approx$ 5 K (7 K before) and allows to run UCN experiments over long-term periods. The somewhat lower temperature of the converter has influence on i) the formation of the sD$_2$ cryo-crystal and ii) UCN losses due to phonon up-scattering which at $T{}\approx$ 5 K equals about the absorption cross section \cite{Ref43, Ref44}.}
\item{Replacement of the horizontal guide (HE4 Nocado\footnote{\url{http://www.nocado.de/}}: $L{}\approx$ 3.8 m) by an also electropolished stainless steel guide but with improved surface quality (HE5 Neumo\footnote{\url{http://www.neumo.de/macroCMS-images/Image/download/Katalog2015-DE.pdf}}). This arrangement improved the transmission by 17 \% as was confirmed in a separate beam time.}
\item{Converter nose and thermal bridge, which were previously coated on their inner surface with NiMo (85/15 weight ratio), were replaced by identical units using $^{58}$NiMo coating. The change of the Fermi potential from $V_\text{F}$= 225 neV to $V_{\text{F}}$ = 311 neV increases the phase space acceptance of UCN produced inside the sD$_2$ crystal, since UCN with $E_{\text{kin}} \le$ (311-105) neV are totally reflected from the wall of the nose, whereas for natural NiMo only UCN with $E_{\text{kin}} \le$ (225-105) neV show total reflection at all angles of incidence.}
\end{enumerate}

\section{UCN density measurements }
\label{sec:2}
\begin{figure}
\centering
\resizebox{0.7\textwidth}{!}{
  \includegraphics{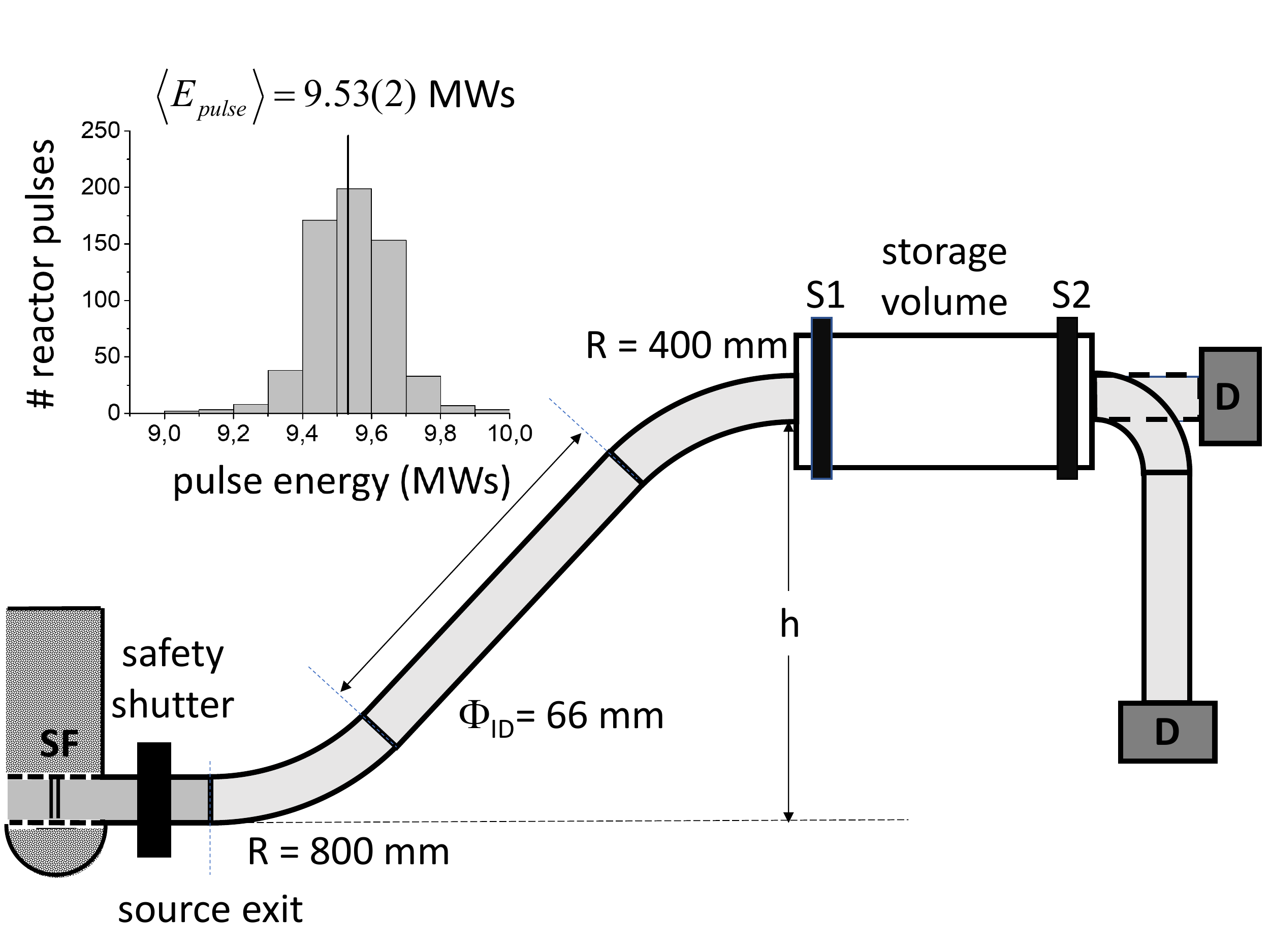}
}
\caption{Sketch of the setup at the TRIGA Mainz (not to scale): Indicated are the positions of the AlMg3 separation foil (SF) on the height of the vertical cryostat and the safety shutter at the exit of the UCN source. Neutrons are guided upwards by means of two 45$^\circ$ bends to a horizontal guide section at height $h$ above the source exit which comprises the actual storage setup. The storage vessel is sketched with shutters S1 and S2, storage volume SV and detector D for both the horizontal and vertical UCN extraction. Inset: Distribution of pulse energies in units MWs.}
\label{fig:2}
\end{figure}

In order to measure the UCN density available at beamport D, we used the storage setup of PSI which was formerly used to perform an experimental comparison of currently operating UCN sources \cite{Ref45}. The 'standard' UCN storage bottle \cite{Ref46} has a volume of $\approx$ 32 litres, comparable in size to typical nEDM experiments. It consists of commercially available electropolished stainless steel tubes ($\diameter_{\text{ID}}$=20 cm) and fast shutters without noticeable losses from openings during the filling and emptying procedures. The cylindrical UCN storage volume of the bottle has a length of 1020.0(11) mm and a resulting volume of  $V_\text{SV}$= 32 044(164) cm$^3$, including the uncertainty on the shutter dimension. The storage bottle was connected to the source exit at beamport D via an S-shaped beamline tube made from two electropolished stainless steel bends (HE4 Nocado) of 800 mm and 400 mm bending radius, respectively, and inner diameter of 66 mm, with straight UCN guides of various lengths in-between (see Fig. \ref{fig:2}). A qualitative information on the UCN energy spectra (soft or hard) is obtained by the ratio between the UCN density measured with the detector in horizontal and vertical extraction positions. The Cascade-U detector \cite{Ref47} based on GEM technology \cite{Ref48} has an Al entrance foil which acts as an energy barrier in horizontal extraction (Al has a reflective wall potential of $V_{\text{F}}$(Al) = 54 neV\footnote{The thickness of native oxides on aluminium alloys and single crystals was investigated in \cite{Ref48a} and found to be $\approx$ 4 nm, i.e., much smaller than the UCN wavelength. Therefore, we believe 54 neV to be a correct assumption for the energy threshold of our Al foils.}). In vertical extraction (detector located 1 m lower), UCN energies are increased by gravity so that some UCN with energies initially below 54 neV can also penetrate the detector foil. Hence, the discrepancy in both detector positions represents the soft part of UCN with energies below 54 neV. The measured UCN densities of the individual neutron pulses are normalised to the average pulse energy $\langle E_{\text{pulse}} \rangle =$ 9.53(2) MWs of the reactor pulses (see inset of Fig. \ref{fig:2}) according to
\begin{equation}
\rho_{\text{i, norm}} = \frac{\text{Counts}_{\text{i}}/V_{\text{SV}}}{E_{\text{pulse, i}}}\cdot{}\langle E_{\text{pulse}} \rangle
\end{equation}

We followed the procedure described in Ref. \cite{Ref45}. A single storage measurement consisted of the following sequence:
\begin{itemize}
\item{\underline{Start:} An electronic trigger signal, coincident with the start of the reactor pulse, starts the timing sequence in the storage experiment electronics.}
\item{\underline{Filling:} The source-side storage bottle shutter (S1) is in the open position to let UCN coming from the source into the bottle. The detector-side storage bottle shutter (S2) stays in the closed position. This state lasts for the defined filling time, $T_{\text{F}}$.}
\item{\underline{Storage:} When the filling time is over, S1 is closed, and both shutters stay closed for the predefined storage time.}
\item{\underline{Counting:} During this period, S1 is closed, while S2 is opened in order to empty the stored UCN into the detector.}
\item{\underline{Trigger Ready:} After the counting time is over, the electronics are reset to the initial state and a new measurement can be started.}
\end{itemize}

\section{Results and discussion}
\label{sec:3}

Storage measurements were performed for:
\begin{enumerate}[label=\Roman*.]
\item{Upgrade a) but still using the in-pile guide of the ``old'' source, i.e., HE4 Nocado tube and NiMo coating of both thermal bridge and nose. For the straight guide section of the S-shaped beamline tube (Fig. \ref{fig:2}), a NiMo-coated quartz tube was used with $L=1.2$ m.}
\item{Upgrade a), b) and c) with the straight UCN guide section replaced by HE5 Neumo tubes  ($L=1.5$ m at the optimum height).}
\end{enumerate}

Setup I: These measurements were exactly the UCN density measurements reported in Ref. \cite{Ref45}. The amount of frozen D$_2$ was 8 mol, whereas 20 mol of H$_2$ were frozen out in the cup-shaped container around the nose (premoderator)\footnote{Freeze-out rate: 1.24 mol/h (H$_2$), 1.04 mol/h (D$_2$). In all cases the premoderator was frozen out before deuterium.} . The same amount of gases was frozen out for Setup II. The data shown in Table \ref{tab:1} refer to measurements with a storage time of 2 s and were obtained after finding the maximal UCN counts by varying both the height $h$ of the installation above the exit of the source (see Fig. \ref{fig:2}) with the maximum at $h$ = 130 cm and the filling times $T_{\text{F}}$. The latter were chosen 4 s for the vertical extraction and 3 s for the horizontal extraction.

\begin{table}
\centering
\caption{Net UCN counts in 2 s storage measurements with the source at Pos. II. Shown are the values from Ref. \cite{Ref45} before the source upgrade, the subtracted UCN leakage counts, and the determined UCN density. The relative amount of soft UCN with $E_{\text{ucn}}\le$ 54 neV can be estimated from the measured horizontal-to-vertical ratio ($h/v$) giving $1-h/v\approx{}0.34$.}
\label{tab:1} 
\begin{tabular}{c c c c}
\hline\noalign{}
Extraction & Net & Subtracted & Density \\
 & UCN counts & leakage counts & (UCN/cm$^3$) \\
\noalign{}\hline\noalign{}
horizontal & 51299(215) & 722(30) & 1.60(1) \\
vertical & 77941(383) & 1229(30) & 2.43(2) \\
\noalign{}\hline
\end{tabular}
\end{table}

The background due to leakage is small ($<$ 2 \%) and the procedure to subtract the UCN leakage counts from the measured UCN counts per storage cycle is discussed in detail in Ref. \cite{Ref45}. \par
\begin{figure}
\centering
\resizebox{0.7\textwidth}{!}{%
  \includegraphics{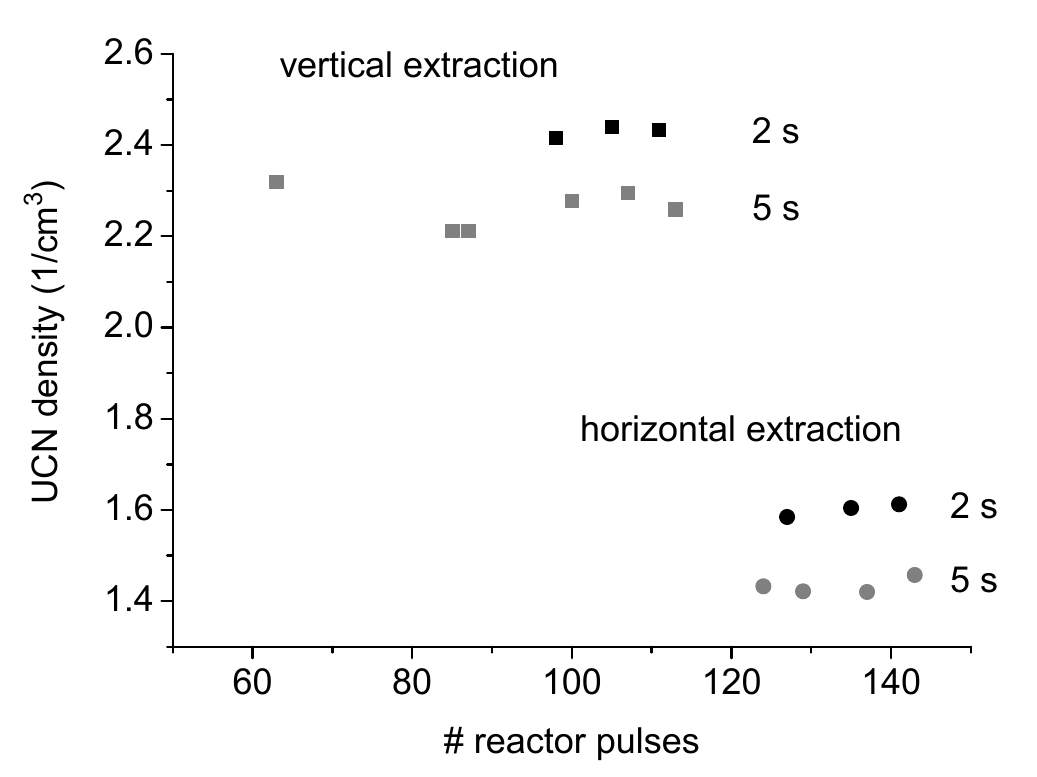}
}
\caption{Measured UCN densities after upgrade a) as a function of the number of reactor pulses. Only data points are shown for storage measurements of 2 s and 5 s both in horizontal and vertical extraction. This data was obtained at Pos. II while the H$_2$ premoderator was prepared and introduced as normal hydrogen. The statistical error bars are smaller than the symbol size.}
\label{fig:3}
\end{figure}

Not discussed, but of relevance in the following is the temporal behaviour of the UCN yield. Fig. \ref{fig:3} shows the measured UCN densities as a function of the number of reactor pulses. Only data points from storage measurements of 2 s and 5 s both for horizontal and vertical extraction are shown (see Table \ref{tab:1} for comparison). The time sequence of data points (apart from the jumps due to the different parameter settings) does not suggest any temporal drift of the UCN yield, i.e., the UCN yield at the average pulse energy of $\approx$ 9.5 MWs is constant at least up to 140  reactor pulses. The same behaviour was observed in former runs before source upgrade a) using hydrogen as premoderator as well (see Fig. 5 of Ref. \cite{Ref40}).

\begin{figure}
\centering
\resizebox{0.7\textwidth}{!}{%
  \includegraphics{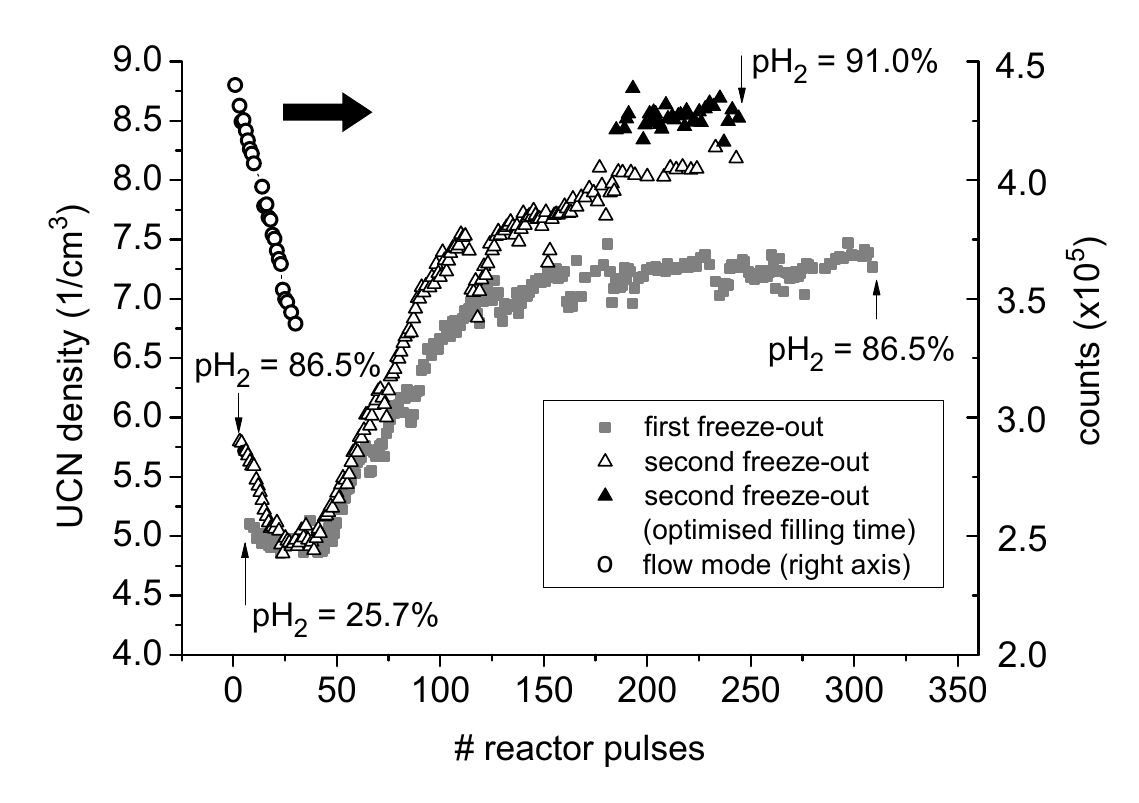} 
}
\caption{UCN densities measured in vertical extraction after upgrade a), b), and c) versus the number of reactor pulses. Two consecutive measurements runs are shown starting with i) normal hydrogen as premoderator (``first freeze-out'') and ii) para-H$_2$/D$_2$ mixture with 5 \% D$_2$ concentration (``second freeze-out''). In the latter measurement run, the source was moved 3 cm closer to the reactor core (Pos. I). The increased thermal neutron fluence is reflected in a $\approx$ 10 \% higher UCN yield. Storage and filling time were set to 2 s and 3.25 s, respectively. From reactor pulse 185 on (``second freeze-out''), the filling times were switched between 3.25 s and 4.5 s with the maximum UCN density achieved at 4.5 s. The para-H$_2$ content was measured at the beginning and at the end of each run. Starting with pure hydrogen (para-H$_2$ 88 \%) as premoderator, the UCN yield decreased by $\approx$ 30 \% already after the first 30 reactor pulses (hollow circles, arrow points to right axis). This data was measured in flow mode, i.e., UCN counts (vertical extraction) with both shutters S1 and S2 open.}
\label{fig:4}
\end{figure}

Setup II: After the full upgrade, i.e., a), b), and c) the situation looks different. Figure \ref{fig:4} shows the UCN densities obtained for storage measurements of 2 s in vertical extraction versus the number of reactor pulses. Two consecutive measurement runs are presented starting with i) normal H$_2$ as premoderator (``first freeze-out'') with the source at Pos. II and ii) para-H$_2$/D$_2$ mixture with 5 \% D$_2$ concentration (``second freeze-out'') with the source moved to Pos. I. In both cases, the storage installation was at $h$ = 142 cm above the exit of the source and the filling time was set to 3.25 s. Only at the end of the second run (from reactor pulse number 185 on), the filling time was alternatively switched between 3.25 s and 4.5 s with the maximum UCN yield obtained at 4.5 s. A total of 35 reactor pulses were made at this last parameter setting from which an average UCN density of 
\begin{equation}
\rho_{\text{ucn}} = 8.53(5) / \text{cm}^3
\end{equation}
could be deduced. The $h/v$-ratio was also extracted, giving $h/v= 0.64(2)$. Within the error bars, it agrees with the $h/v$-ratio measured before (see Table \ref{tab:1}), i.e., we see the same UCN fraction above and below the Al threshold energy (data summarised in Table \ref{tab:2}).

\begin{table}
\centering
\caption{Net UCN counts in 2 s storage measurements after the source upgrade. The first row shows the average values (vertical extraction) of the net UCN counts and the resulting UCN density from storage measurements between reactor pulse nos. 185 and 220 with the source at Pos. I and filling time set to 4.5 s. The horizontal-to-vertical ratio ($h/v$) was measured at Pos. II of the source (2$^{\text{nd}}$ and 3$^{\text{rd}}$ row) giving $(h/v) = 0.64(2)$.}
\label{tab:2}
{\begin{tabular}{c c c c c c}
\hline\noalign{}
Extraction & Net & Subtracted & Density & Filling & h/v  \\
 & UCN counts & leakage counts & (UCN/cm$^3$) & time (s) \\
\noalign{}\hline\noalign{}
vertical/& 273281(483) & 2723(55) & 8.53(5) & 4.5 & \\
Pos. I  & & & & & \\
horiz./ & 159900(2860) & 1778(45) & 4.98(10) & 2.5 & \\
Pos. II & & & & & \\
vertical/ & 250340(3050) & 2686(55) & 7.81(10) & 3.0 & 0.64(2) \\
Pos. II & & & & & \\
\noalign{}\hline
\end{tabular}}
\end{table}

\begin{figure}
\centering
\resizebox{0.7\textwidth}{!}{
  \includegraphics{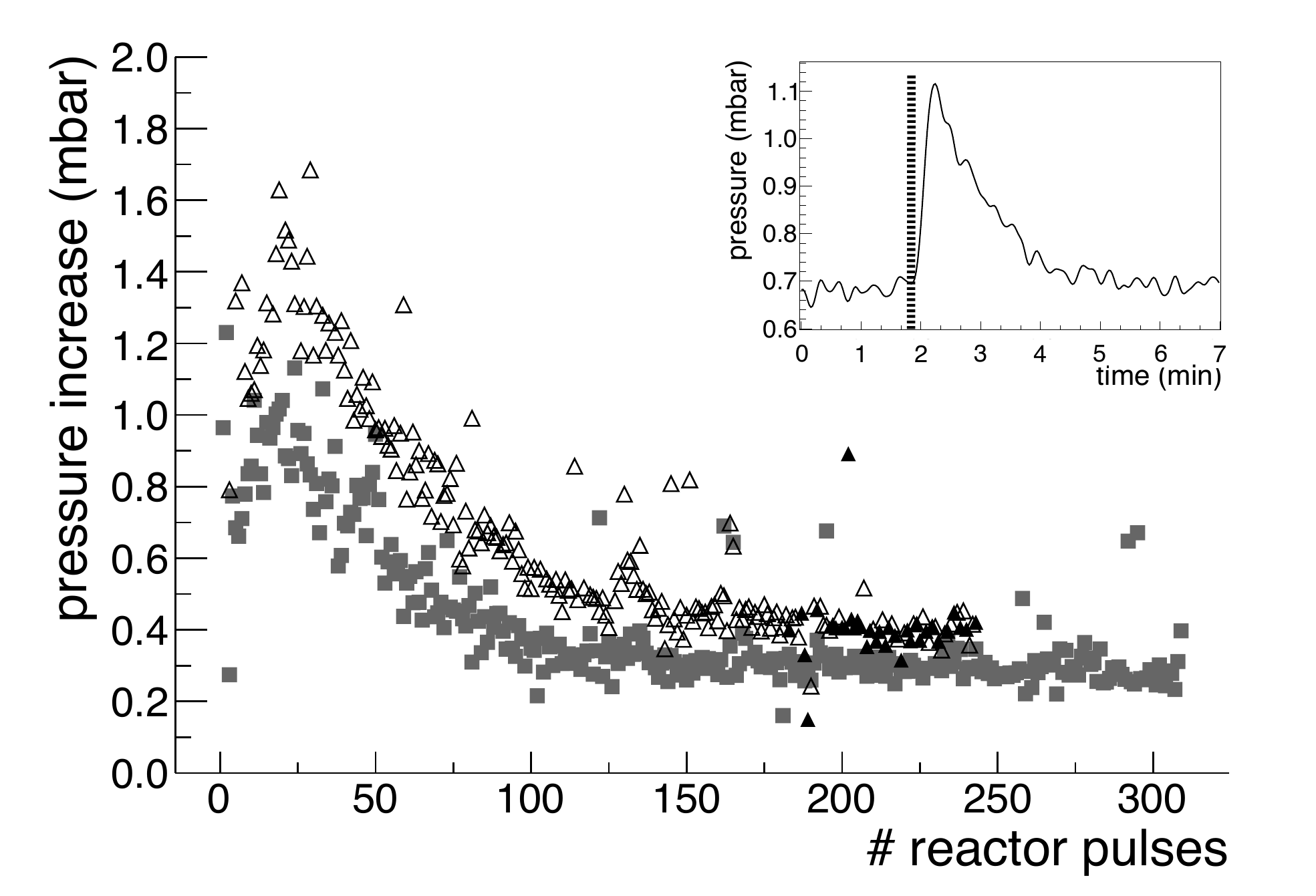}
}
\caption{Measured maximum pressure increase in the horizontal guide section versus the number of reactor pulses. Shown again are the results of the ``first freeze-out'' (grey squares) and the ``second freeze-out'' (triangles), which allows for the identification of some correlations with the measured UCN densities of Fig. \ref{fig:4} for the two consecutive measurement runs (see text for details). Inset: Deuterium pressure measured after a reactor pulse (indicated by the dashed vertical bar). The pressure decrease after the peak ($p_0$) is due to recondensation. The pressure value of $\approx$ 0.7 mbar is the sensor offset.}
\label{fig:5}
\end{figure}

The general finding is that the measured UCN densities have increased in both cases (see Table \ref{tab:1} for comparison) and the UCN yield follows the same pattern, i.e., after an initial decrease, a plateau builds up in the UCN yield between pulse number 10 and 40 with a subsequent increase leading to a saturation value about 40 \% higher than the initial UCN yield. In these measurements, the ratio of ortho-H$_2$ to para-H$_2$ was determined offline using the Raman spectrometer. This was done just before freeze-out and immediately after the run. Starting with normal hydrogen ($\approx$ 25 \% para-H$_2$), the conversion into para-H$_2$ obviously happens within $\approx$ 50 h ($\approx$ 250/$R_{\text{pulse}}$). Since the expected ortho-to-para conversion time in solid hydrogen is approximately 800 h \cite{Ref49}, the formation of radicals in solid hydrogen due to the intense neutron and gamma exposure during the reactor pulses probably speeds up the ortho-para conversion process \cite{Ref50}. Para-H$_2$ is an attractive medium for neutron moderation because of the inelastic $J=0$ to $J=1$ transition between the rotational energy levels of the molecule (thereby inducing a para-to-ortho transition); these levels are separated by 14.7 meV and the partial cross sections for such an event is largely above 14 meV \cite{Ref51, Ref52, Ref53}. This makes para-H$_2$ a useful material to increase the production of cold neutrons, as elastic scattering\footnote{Elastic scattering refers to the centre-of-mass frame. If the effective mass of the target, i.e., the entire lattice, is very large, the neutron does not lose energy in the scattering process.} is less efficient in moderating thermal neutrons from the reactor.  So, at first glance, the spectrum of incoming neutrons is better adapted to the phonon energy spectrum of the sD$_2$ converter \cite{Ref43, Ref54} which again would result in higher UCN densities and would explain, at least qualitatively, the measured increase of the UCN yield towards a saturation value. This, however, is in contradiction to the temporally constant UCN yield observed so far (see Fig. 2 and Fig. 5 in Ref. \cite{Ref40}, too). Or to put it another way, the expected increase in the UCN yield may be masked by other competing processes. Probably the most relevant change in the upgrade was the use of $^{58}$Ni instead of natural Ni for the NiMo coating of the converter cup and the thermal bridge. How this translates into the observed temporal dependence of the UCN yield (Fig. \ref{fig:4}) is difficult to predict, since parameters like temperature gradients, crystal quality, heat load and their mutual dependencies are not sufficiently known. \par
The recorded D$_2$ pressure curves are a strong indication that the thermal properties of the converter due to the evolution of its crystalline structure and spatial distribution are reflected in the measured UCN yield. The increase of the deuterium pressure in the horizontal guide ($V$ = 12.8 L) shortly after the reactor pulse was monitored during the curves presented in Fig. \ref{fig:4}. The monitoring of the pressure was performed using a pressure transducer close to the AlMg3 separation foil at the end of the guide section. The maximum pressure increase $p_0$ (see inset of Fig. \ref{fig:5}) was investigated to study the influence of heat load on the performance of the sD$_2$ converter. What can be stated first is that the source at Pos. I (3 cm closer to the reactor core) shows overall increased pressure values due to the higher deposited energy (Fig. \ref{fig:5}). The general course of the recorded pressure data, however, follows the same pattern: An increase of $p­_0$ during the first reactor pulses with a maximum between pulse number 10 and 40 can be identified, which is followed by a steady decrease settling at a minimum $p_0$ value from reactor pulse $\approx$ 100 on. Therefore, from the general course of the $p_0$ curves after averaging over pulse-to-pulse fluctuations, the following can be derived: The measured UCN yields in Fig. \ref{fig:4} show an almost reciprocal behaviour indicating that the converter's thermal properties largely reflect the temporal behaviour of the UCN yield. \par
A better insight into the role of the H$_2$ premoderator with regard to the subsequent freezing process of the sD$_2$ crystal and the measured UCN yield is obtained from another observation made:  Freeze-out of para-H$_2$ (88 \%) as premoderator instead of normal H$_2$ showed that the UCN yield decreased by $\approx$ 30 \% already after the first 30 reactor pulses (see Fig. \ref{fig:4}). Thus, a completely different temporal behaviour of the UCN yield was obtained. 
The premoderator around the actual converter cup has a strong impact on the temperature distribution on the walls of the nose as well as on the directly adjacent parts of the neutron guide (thermal bridge). In going from normal to para-H$_2$, the thermal conductivity of solid hydrogen increases dramatically ($>$ 10) \cite{Ref55} . Therefore, it is no surprise that depending on the respective ortho/para ratio of the H$_2$ premoderator, the subsequent freeze-out of D$_2$ from the gas phase results in different shapes of the cryo-crystal within the converter volume. Moreover, during pulsation, the heat load leads to a re-shaping of the crystal that can be more or less pronounced depending on the original formation of the crystal shortly after freezing\footnote{Investigations performed at the PULSTAR UCN source \cite{Ref36}: By means of optical elements, the growth, formation, and re-shaping of sD$_2$ could be visualised in-situ under different temperature conditions and thermal treatments in the converter cup \cite{Ref56}, which underpins our observations.}. In Fig. 3 of Ref. \cite{Ref57}, the change in thermal conductivity of (para-H$_2)_{1-c}$ (D$_2)_c$ versus the D$_2$ concentrations ($c$) in the temperature range $2 \; \text{K} < T < 9 \; \text{K}$ is shown. Already a D$_2$ concentration of $c = 0.05$ reduces the thermal conductivity by a factor of 10 and values comparable to normal hydrogen at the same temperature are obtained. The measured UCN yield in Fig. \ref{fig:4} with the para-H$_2$/D$_2$ premoderator at 5 \% deuterium concentration not only shows the same temporal dependence, even the absolute numbers agree with each other when taking into account that this measurement was conducted at Pos. I, i.e., at a slightly higher thermal neutron flux resulting in a $\approx$ 10 \% higher UCN yield. The UCN density difference for the measurements with 25.7 \% and 86.5 \% para-H$_2$ obviously is rather small, given the large concentration difference. This finding hints at a rather small influence of para-H$_2$ on the total UCN rate. \par
The reproducibility of these measurements could also be confirmed in separate runs. Typically after $\approx$ 100 reactor pulses, the saturation value of the UCN yield is reached and remaining fluctuations are then less than 3 \%. This relative stability of the UCN yield could be observed over the measurement period of 2 weeks.   

\section{Storage curve measurement and Monte Carlo simulation}
\label{sec:4}

\begin{figure}
\centering
\resizebox{0.7\textwidth}{!}{%
  \includegraphics{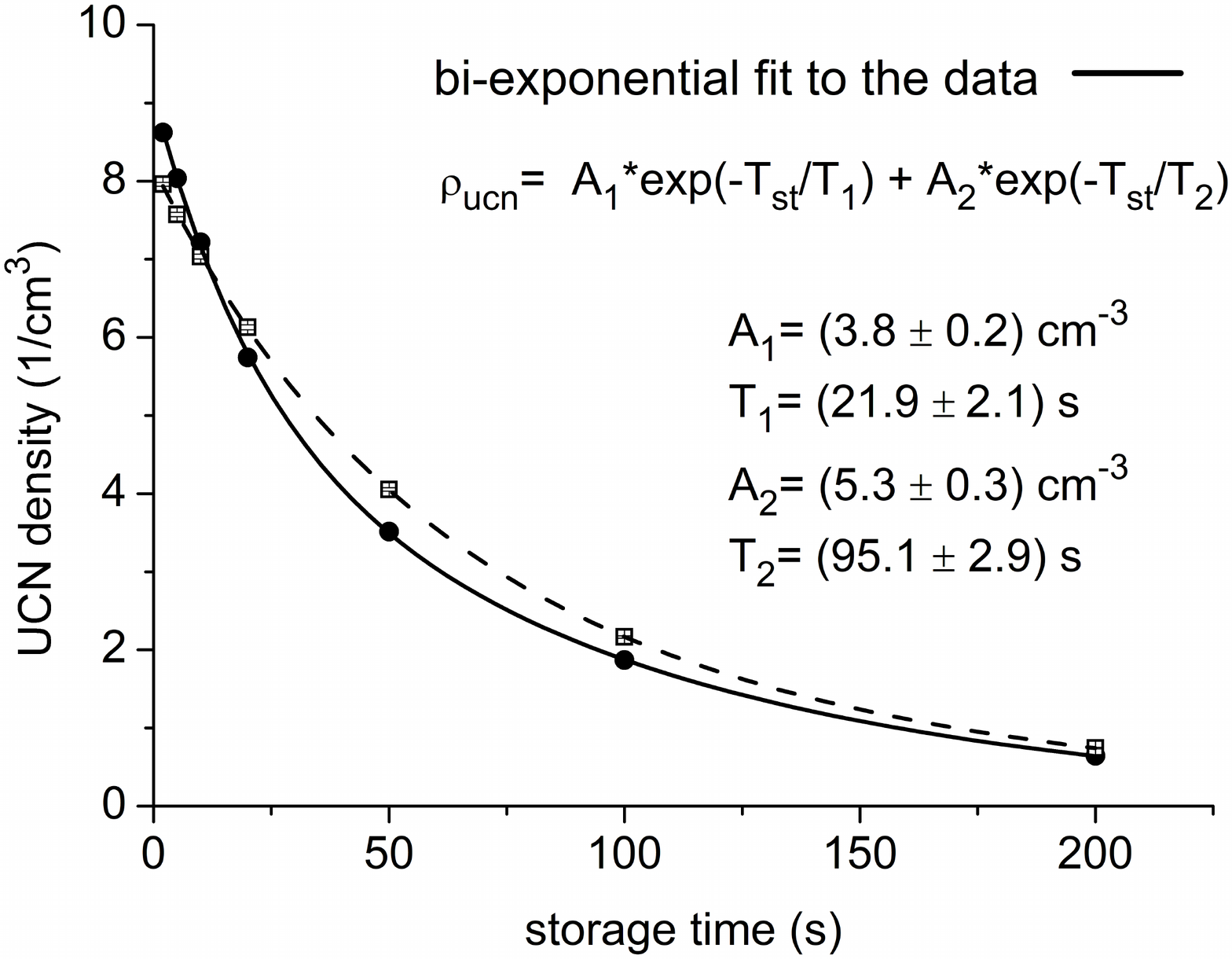}
}
\caption{UCN density as a function of the storage time $T_\text{st}$ in a volume of $V{}\approx$ 32 L (vertical extraction). The error bars are smaller than the symbol size. The filling time was set to 4.5 s. The functional dependence of $\rho_{\text{ucn}}(T_{\text{st}})$ can be described using a bi-exponential function (solid line). The storage curve obtained by a custom-made MC transport simulation (dashed line) generally reproduces the measured data in a reasonable approximation:  $A_1^{\text{sim}}= (4.2\pm{}0.6)$ cm$^{-3}$, $T_1^{\text{sim}}= (47.2\pm{}3.2 )$ s , $A_2^{\text{sim}}= (4.1\pm{}0.6)$ cm$^{-3}$ , and $T_2^{\text{sim}} = (113.4\pm{}6.8)$ s. An apparent discrepancy only shows up in the simulation of the fast time component ($T_1$) of the storage curve which is predicted to be almost by a factor of 2 larger. We attribute this discrepancy to the insufficient knowledge of storage parameters for the hard part of the UCN energy spectrum.}
\label{fig:6}
\end{figure}

After having found the parameter settings for the maximum UCN density in the `standard' storage volume of $\approx$ 32 L, the storage curve was measured. The results are shown in Fig. \ref{fig:6}, in which the respective UCN densities are plotted for predefined storage times $T_\text{st}$ in the range $2 \; \text{s} \le T_{\text{st}} \le 200 \; \text{s}$. A bi-exponential fit to the data points describes the functional dependence best.

The expected UCN densities from a Monte Carlo (MC) simulation are also shown in Fig. \ref{fig:6}. The source is modelled with a MC transport code by setting starting point, energy and direction for neutrons in the sD$_2$ converter (assuming neutrons with a normalised Maxwellian tail velocity distribution ($\sim{}v^2$) \cite{Ref43}) and transporting them along the guide to the storage vessel and finally towards the detector. Details on the MC simulation of the UCN D source performance can be found in Ref. \cite{Ref40}. The simulated storage curve agrees fairly well with the measured one and confirms the functional dependency. It should be noted that the MC data in absolute numbers are not scaled, but are derived directly from the one-phonon down-scattering cross section in the incoherent approximation \cite{Ref43} using the analytical form of the phonon density spectrum of solid ortho-deuterium derived from neutron inelastic scattering \cite{Ref58}. \par
A more refined analysis of the UCN production cross section by \cite{Ref54} incorporates coherent contributions and multi-phonon excitations leading to a $\approx$ 20 \% higher value of the down-scattering cross section.The reason of using the simplified approach is our limited knowledge of the parameters governing  the UCN transport which gives us an overall uncertainty of $\approx$ 30 \% in the total number of predicted UCN. In order to produce a good fit to the measured data, we allowed transport parameters like loss coefficient, diffuse reflection probability, as well as the parameters which govern the transparency of the sD$_2$ crystal to vary within some narrow range. \par
For the mean free loss length ($\lambda_{\text{loss}}$) of UCN in the sD$_2$ converter, we used the values derived in [40] resulting in 4.6 cm $< \lambda_{\text{loss}}$ $<$ 10.0 cm at $T{}\approx$ 5 K. For the MC simulation, the sD$_2$ crystal has the form of a spherical meniscus with curvature radius $R_c$ = 3.5 cm and length $L$ = 6.2 cm. Losses of UCN due to up-scattering in the D$_2$ gas can be neglected, since most of the UCN have traversed the horizontal guide ($\leq$ 1 s) before the evolution of the D$_2$ gas pressure sets in (see inset of Fig. \ref{fig:5}). \par
The material-dependent loss coefficient  $\eta= W/V$, defined as the ratio of imaginary ($W$) and real part ($V$) of the Fermi potential $V_{\text{F}} = V-iW$ \cite{Ref2}, was taken to be $\eta = 3\times{}10^{-4}$ to describe UCN storage in the PSI storage vessel\footnote{The stainless steel storage bottle has a calculated Fermi potential on the surface of 185 neV and 40 \% of diffuse scattering on the surface. An almost equal number for $\eta$ was used in the PSI developed simulation tool MCUCN \cite{Ref59} to describe their measured storage curves and UCN energy spectra with the `standard' storage bottle \cite{Ref45}.}. \par
\begin{figure}
\centering
\resizebox{0.7\textwidth}{!}{%
  \includegraphics{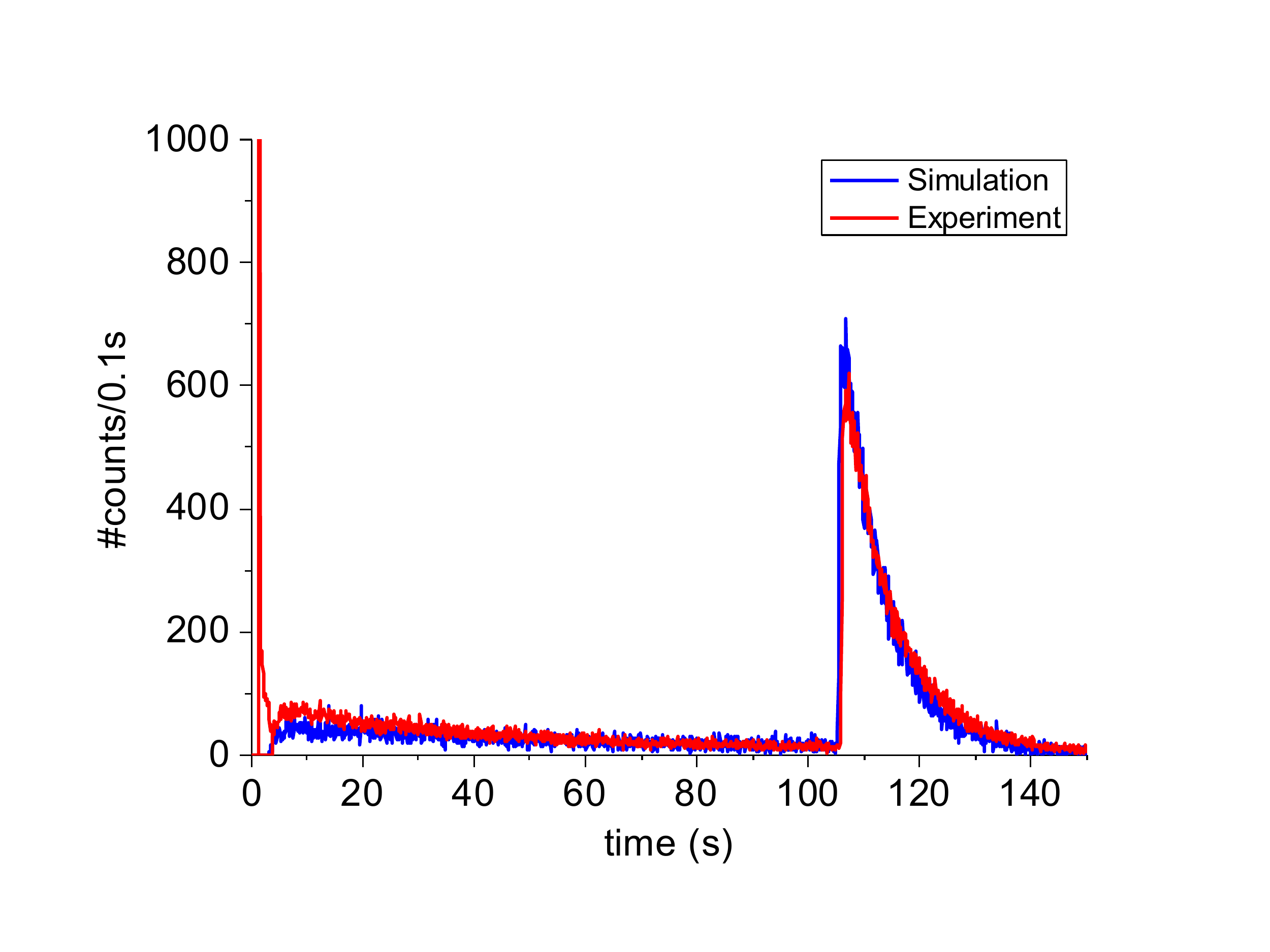}
}
\caption{Time spectrum of UCN counts per 0.1 s (red data points) with vertical extraction. At $t=0$, the reactor pulse signal (thermal neutron peak) is seen. It follows the filling time of 4.5 s after the UCN storage period of 100 s and finally the extraction. The MC simulation (blue data points) describes both the UCN leakage through shutter S2 during storage and the emptying peak from t $\approx$ 105 s on when S2 was opened.}
\label{fig:7}
\end{figure}

The time spectrum of UCN arrival at the detector is shown in Fig. \ref{fig:7} for a storage time of 100 s. UCN were leaking through shutter S2 during the period of storage, which could be observed. Then, shutter S2 was opened and the UCN remaining in the storage vessel were detected. The MC simulation does reproduce the experimental data. 

\begin{figure}
\centering
\resizebox{0.7\textwidth}{!}{%
  \includegraphics{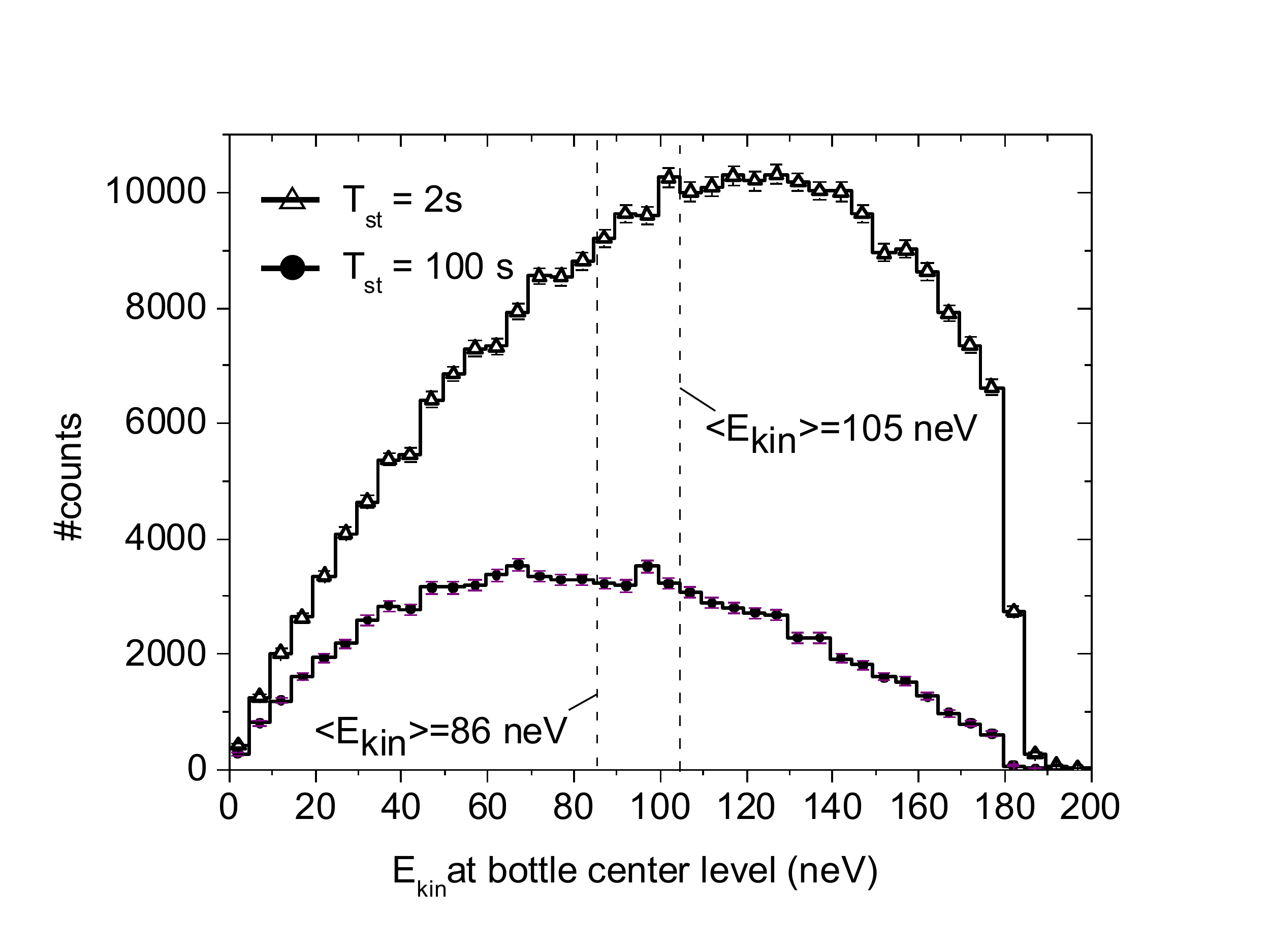}
}
\caption{Simulated UCN energy spectra in the storage bottle after a storage time of 2 s and 100 s. The dashed lines reflect the different UCN mean energies of 105 neV and 86 neV for the two energy distributions. The difference of the two energy spectra is due to the energy-dependent loss cross sections, which result in a faster loss of neutrons with higher energies.}
\label{fig:8}
\end{figure}

Figure \ref{fig:8} shows the simulated time-dependent UCN energy spectra for storage times of 2 s and 100 s. The UCN mean energies decrease for later times in which a softer UCN spectrum is present in the storage bottle. The MC simulation gave $\approx$ 10 \% losses due to the AlMg3 entrance foil of the detector for vertical extraction and $\approx$ 45 \% in case of horizontal UCN extraction. Similar numbers were obtained in foil transmission measurements \cite{Ref60}.

\begin{figure}
\centering
\resizebox{0.7\textwidth}{!}{%
  \includegraphics{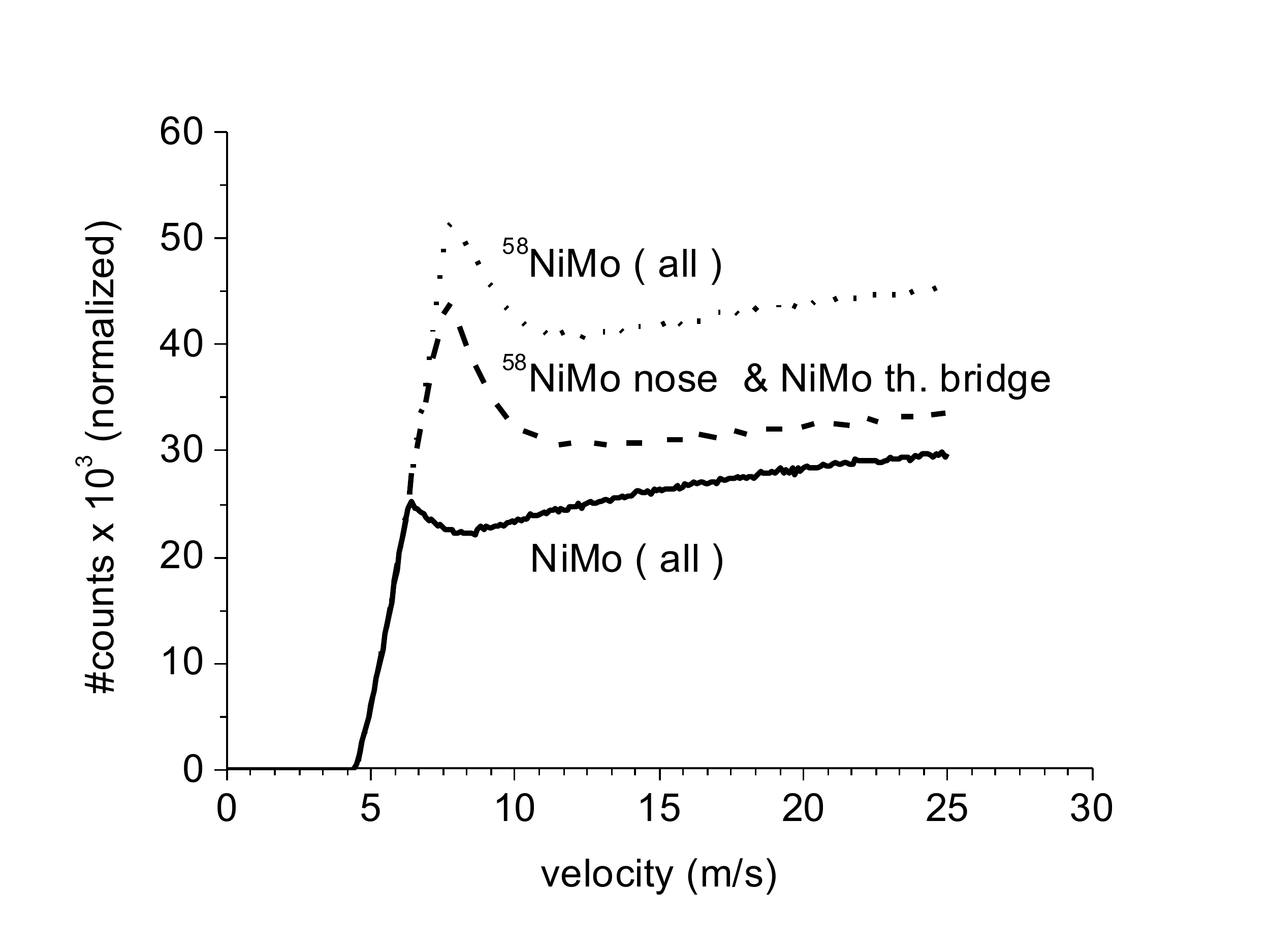}
}
\caption{Simulated UCN/VCN velocity spectrum at the exit of the thermal bridge for different wall coatings at the front part of the horizontal guide including the nose: a) NiMo coating of nose \& thermal bridge, b) Ni replaced by $^{58}$Ni at the nose, and c) all $^{58}$NiMo-coated. See text for details.}
\label{fig:9}
\end{figure}

Finally, the measured gain factor in the UCN density of 3.5 in going from Setup I to Setup II can be revealed from a MC simulation of the transient UCN/VCN velocity spectrum\footnote{VCN are very cold neutrons, i.e., non-storable neutrons with velocities $v > v_c$, where $v_c$ is the critical velocity set by the material's potential.} downstream at the exit of the thermal bridge. In Fig. \ref{fig:9}, the corresponding spectra are shown  for i) entirely NiMo-coated nose \& thermal bridge (solid line), ii) Ni replaced by $^{58}$Ni at the nose (dashed line) as an intermediate step, and iii) all $^{58}$NiMo-coated (dotted line). The onset is at $\approx$ 4 m/s in all cases with a steep rise up to a maximum value of counts at $v\approx{}v_c = \sqrt{2V_{\text{F}}(\text{NiMo})/m_n}$, which is about  6.6 m/s for NiMo and 7.8 m/s for the nose coated with $^{58}$NiMo. After a slight drop, an increase is observed again which continues up to the maximum simulated UCN/VCN velocity range of 25 m/s. Below 10 m/s, the dominant gain in UCN yield comes from the $^{58}$NiMo coating of the nose itself which houses the sD$_2$ converter crystal, whereas the higher Fermi potential of the $^{58}$NiMo-coated thermal bridge provides a larger phase-space acceptance in particular for VCN which exit the nose downstream. All in all, a gain factor of $\approx$ 2.2 is to be expected for UCN/VCN with $v < 10$ m/s. This number further increases taking the improved UCN/VCN transmission through the entire remaining neutron guide of Setup II into account. As mentioned above, there is a 17 \% increase in transmission due to the replacement of the 3.8 m long horizontal guide by HE5 (Neumo) tubes. An additional improvement of 15 \% was obtained by replacing the straight guide section of the external beam line (NiMo-coated quartz tubes as used in Setup I) with HE5 (Neumo) tubes.  Summarised, one expects a gain factor of $\approx$ 3 which is close to the measured increase of the UCN yield for Setup II. With the MC simulation to model the source, a deeper insight into the UCN production and loss mechanisms is accessible which helps to further improve the source's performance. With the present upgrade, a powerful and reliable UCN source at the pulsed reactor TRIGA Mainz is put into operation. Within the framework of the cluster of excellence PRISMA, the neutron lifetime experiment $\tau$SPECT with 3D magnetic storage will be performed using this source.

\section{Summary}
\label{summary}

With the described and conducted measures as part of our source upgrade, the UCN densities at the pulsed reactor TRIGA Mainz could be improved by a factor of 3.5 to $\rho_{\text{ucn}} \approx 8.5 /$ cm$^3$. The measurements were performed with the “standard” UCN storage bottle of $V = 32$ L which has recently served for an experimental comparison of the leading UCN sources including the UCN source at TRIGA Mainz. MC simulations of the full transport of UCN from the production location until detection reproduce the measured storage curve and the ratio ($h/v$) of the UCN yield in horizontal and vertical extraction. After the sequential freeze-out of the premoderator (H$_2$) and the converter (sD$_2$), the time course of the UCN yield in pulse mode always follows the same pattern finally reaching its saturation value after about 100 reactor pulses. From there on, the fluctuation is less than 3 \% over measurement periods of 2 weeks.

\section{Acknowledgment}
\label{acknowledgment}

We would like to thank the staff of the reactor and the workshops of the Institute of Nuclear Chemistry and the Institute of Physics for their help during the source upgrade and in setting up the He liquefier as well as their assistance during the experiments. This work was supported by the cluster of excellence PRISMA ``Precision Physics, Fundamental Interactions and Structure of Matter'', Exc 1098 and by the DFG Graduate School ``Symmetry Breaking in Fundamental Interactions''. The PSI acknowledges the support of the Swiss National Science Foundation under Projects 200020\_149813 and 200020\_163413.

\end{document}